\documentclass{SLC}
\articlenumber{4}
\graphicspath{{.}{../}{../Figures/}{../../}}

\usepackage{amsfonts,amsmath,amssymb}

\def\be{\begin{equation}}
\def\ee{\end{equation}}
\def\bea{\begin{eqnarray}}
\def\eea{\end{eqnarray}}
\def\tr{{\rm tr}\,}
\def\bfn{{N}}
\def\Tr{\textbf{Tr}\;}

\def\be{\begin{equation}}
\def\ee{\end{equation}}

\def\R{\mathbb{R}}
\def\Q{\mathrm{Q}}
\def\rme{\mathrm{e}}
\def\rmi{\mathrm{i}}

\author[S.~Ouvry]{St\'ephane Ouvry\textsuperscript{1}\protect\orcid{0000-0003-4258-9506}}
\author[A. P.~Polychronakos]{Alexios P. Polychronakos\textsuperscript{2}\protect\orcid{0000-0002-0443-2797}}
\address{\textsuperscript{1}LPTMS, CNRS, Universit\'e Paris-Saclay, France;\newline \hspace*{4.5mm} \rm \url{http://lptms.u-psud.fr/fr/blog/stphane-ouvry}}
\address{\textsuperscript{2}Physics Department, City College of New York, USA;\newline 
\hspace*{4.5mm} Graduate Center, City University of New York, USA;\newline 
\hspace*{4.5mm} \rm \url{https://www.ccny.cuny.edu/profiles/alexios-polychronakos}}

\title[Signed area enumeration for lattice walks]{Signed area enumeration for lattice walks}
\abstract{We give a summary of recent progress on the signed area enumeration of closed walks on planar lattices. Several connections are made with quantum mechanics and statistical mechanics. Explicit combinatorial formulae are proposed which rely on sums labelled by the multicompositions of the length of the walks.}
\keywords{lattice walks, signed area enumeration, Hofstadter model, exclusion statistics}
 
\begin{document}
\maketitle
%\href{mailto:stephane.ouvry@universite-paris-saclay.fr}{stephane.ouvry@universite-paris-saclay.fr}, \href{mailto:apolychronakos@ccny.cuny.edu}{apolychronakos@ccny.cuny.edu}

\section{Introduction}

The seminal problem of the signed area enumeration of walks on planar lattices of various kinds has been around for a long time. It is well known that this purely combinatorial problem can be equivalently reformulated in the realm of Hofstadter-like quantum mechanics models
(note that, in physics, the ``signed area'' is often called the ``algebraic area''). Recently, in~\cite{1},
 this problem has been given a boost in the form of an explicit enumeration formula which in turn could be reinterpreted~\cite{2,2B,2C} in terms of statistical mechanics models with exclusion statistics, again a purely quantum concept. It is a striking fact that an enumeration quest regarding classical random walks should be in the end so intimately connected to quantum physics. 

In this note we give a summary of this recent progress starting with the original signed area enumeration problem for closed walks on a square lattice and then enlarging the perspective to other kinds of lattices and walks via the statistical mechanics reinterpretation. The first question we address is: Among the $\binom{\bfn}{{\bfn}/2}^2$ closed $\bfn$-step walks that one can draw on a square lattice starting from and returning to a given point (note that ${\bfn}$ is then necessarily even), how many of them enclose a given signed area $A$?
		
The signed area enclosed by a {directed} walk is weighted by its {\it winding number}\/: If the walk moves around a region in a counterclockwise direction, its area counts as positive, otherwise it counts as negative; if the walk winds around more than once, the area is counted with multiplicity. These regions inside the walk are called winding sectors.
\pagebreak

\begin{figure}
\centering
	\includegraphics[width=.8\textwidth]{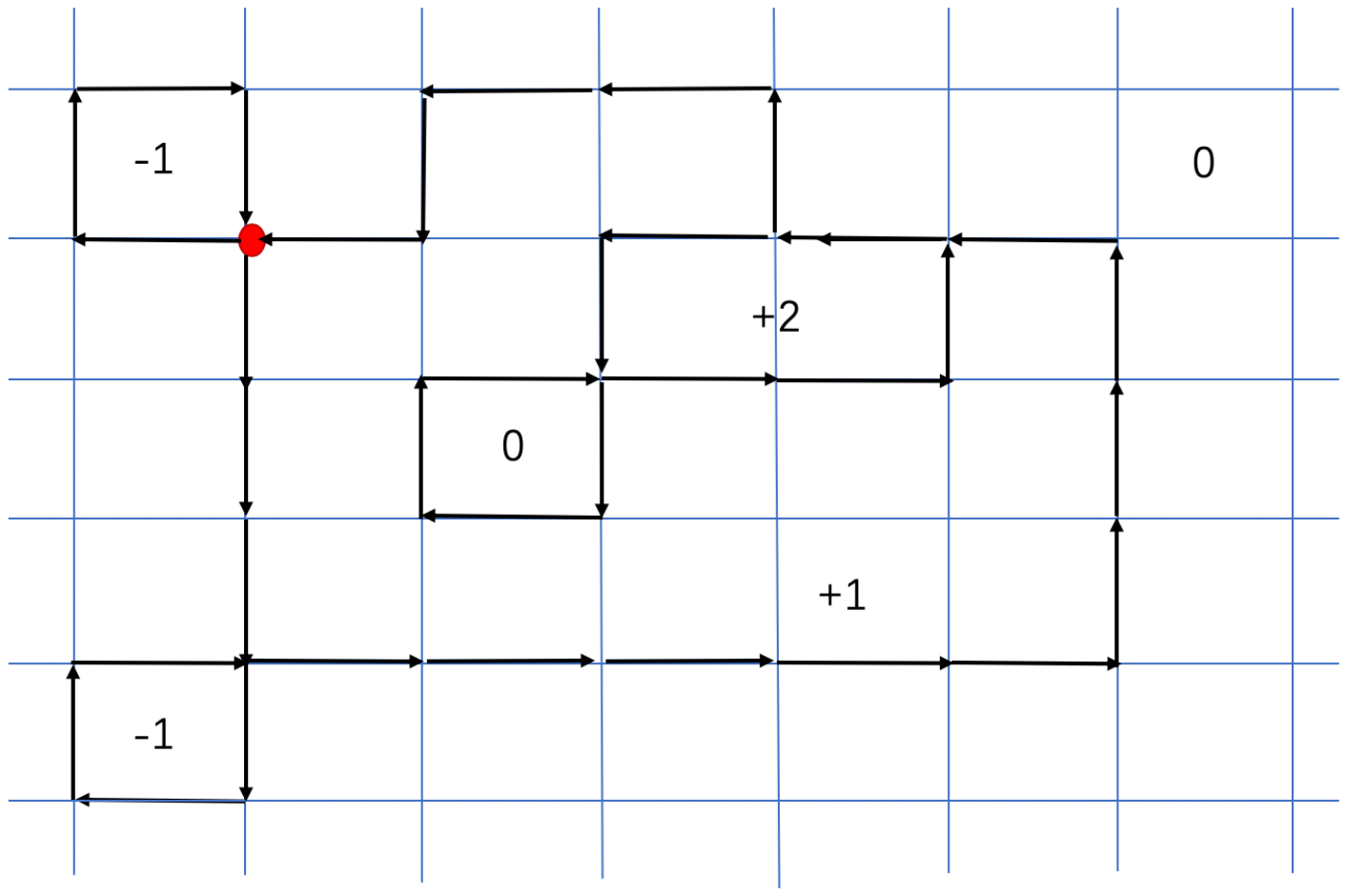}
\setlength\abovecaptionskip{-10mm}
	\caption{A closed walk of length ${\bfn}=36$ starting from and returning to the same bullet red point with its various winding sectors $m=+2,+1,0,-1,-1$
(containing respectively $2,14,1,1,1$ unit lattice cells).
Note the double arrow on the horizontal link (above the +2 sector) which indicates that the walk has moved twice on this link, here in the same left direction.}\label{fig1}
\end{figure}
\smallskip
In Figure~\ref{fig1}, the $0$-winding sector inside the walk %(to be distinguished from the infinite $0$-winding sector outside the walk) 
arises from a superposition of a $+1$ and a $-1$ winding. %It does not contribute to the signed area. 
Summing the areas of each sector, with the corresponding multiplicative weight, 
gives the signed area $A={ (-1)}\times { 2}+(+ 0) \times 1 + {(+1)}\times { 14}+{ (+2)}\times { 2} =16.$

\def\x{\mathbf{x}}
More formally, if $\gamma: [0, 1] \rightarrow \R^2$ is a closed path that begins and ends at the origin,
the signed area of this path is 
\be A = \int_{\R^2} \eta(\gamma,\x)d\x= \sum_{m=-\infty}^{\infty}m S_m,\ee
where $\eta(\gamma,\x)$ is the winding number of $\gamma$ around the point $\x\in \R^2$,
and where $S_m$ denotes the classical area of the $m$-winding sectors inside the path (i.e.~the number of unit lattice cells it encloses with winding number $m$, where $m$ can be positive or negative).

%See e.g.~\cite{Lee2021}.
%More generally, calling $S_m$ the classical area of the $m$-winding sectors inside a walk (i.e.~the number of unit lattice cells it encloses with winding number $m$, where $m$ can be positive or %negative) the signed area is
%to be distinguished from the classical area 
%$\sum_{m=-\infty}^{\infty} S_m$.

Winding sectors for continuous Brownian curves as well as for discrete lattice walks have been the subject of study for a long time. In this respect, we note in the last few years some advances in~\cite{titi} 
 where\footnote{Note that Timothy Budd gave a talk on his article~\cite{titi} at the conference~\href{https://lipn.univ-paris13.fr/~cb/LPC/2021/Photos/}{Lattice Paths, Combinatorics and Interactions}, at CIRM, in 2021.
A video of his talk is available on the website of this conference.} an explicit formula for the expected area $\langle S_m\rangle$ of the $m$-winding sectors inside square-lattice walks is proposed, to the exception of the $0$-winding sector, for the simple reason that the latter is difficult to distinguish from the outside (i.e.~$0$-winding again) sector, which is of infinite size. Taking the continuous limit allows us to recover the results previously obtained in~\cite{nous} for Brownian curves. One notes that for Brownian curves the expected area $\langle S_0\rangle$ of the $0$-winding sectors is also known by other means thanks to the SLE machinery~\cite{Tru}. However, it remains an open problem for discrete lattice walks.
\pagebreak

{{Counting the number of closed walks of length $\bfn$ {on the square lattice} enclosing a signed area $A$ {can be achieved in a} most straightforward way by introducing two lattice hopping operators (symbols) $u$ and $v$ 
respectively in the right and up directions, {as well as operators $u^{-1}$ and $v^{-1}$ corresponding to hops in the left and down direction. A directed walk on the square lattice starting at the origin
is then represented by the
ordered product of the hopping operators corresponding to its individual steps. By convention we order
the operators \textit{from right to left} as we trace the steps of the walk: $vu$ corresponds
to a up step $u$, followed by a right step $v$. Clearly the set of all walks
of length $\bfn$ on the lattice is reproduced by the $4^{\bfn}$ terms in the expansion of
\bgroup
\setlength{\abovedisplayskip}{4.75pt}
\setlength{\belowdisplayskip}{4.75pt}
\begin{equation}
\big(u+u^{-1}+v+v^{-1}\big)^{\bfn}% := H^{\bfn}
\label{Hn}
\end{equation}
into monomials of products of symbols, each with $\bfn$ factors. The operator $
 u+u^{-1}+v+v^{-1}$
can be considered as the generator of walks.

We are interested in closed walks, and in counting their multiplicity according to their signed area.
To this end, we endow the above operators with the relations
\be
u u^{-1} = u^{-1} u = v v^{-1} = v^{-1} v = 1,
\label{noncom}
\ee
to which we add a non-commutativity relation (which expresses the fact that the elementary walk circling one lattice cell in the counterclockwise direction has signed~area~$1$):
\be
 v^{-1} \; u^{-1} \; v \; u =\Q,
\ee
where $\Q$ is a central element (that is, $\Q$ commutes with all operators).
This entails 
\begin{equation}\label{add_rel}
v u=\Q u v, \quad vu^{-1}=\Q^{-1} u^{-1} v, \quad v^{-1} u=\Q^{-1} u v^{-1}, \quad v^{-1} u^{-1} =\Q u^{-1} v^{-1},
\end{equation}
\egroup
which allows us to reduce all terms in~\eqref{Hn} into monomials of the form $u^m v^n$, $(m,n)\in{\mathbb Z}^2$ being the lattice
coordinates of the end of the walk, with coefficients which are powers~of~$\Q$. In particular, closed walks
correspond to the monomial $u^0 v^0$ in~\eqref{Hn}.

The non-commutativity relation $v u=\Q\; u v$ has the effect of flipping a right-up two-step segment 
into an up-right one, producing a factor of $\Q$, and similarly for each relation~in~\eqref{add_rel}. In each case, the coefficient is
$\Q$ to the power of the signed area 
of the unit lattice cells left behind by the exchange. Repeated application
of these relations reduces each walk to hook-shape walk $u^m v^n$ with an overall coefficient
$\Q^A$,
with $A$ 
% of the total signed area of the original walk closed by joining its end to the origin with a vertical and a horizontal straight walk (in that order). 
the signed area of the original walk prolonged into a closed walk by joining its end to the origin with a vertical and a horizontal straight walk.
In particular, closed walks correspond~to~monomials~$\Q^A u^0 v^0$, with $A$ the signed area of the walk. 
This area is maximal if the walk forms a square of width $N/4$, and then the signed area is $\pm N^2/16$.
Therefore, using the notation $[u^0v^0]$ for the extraction of the constant term in a Laurent polynomial in $u$ and $v$,
the distribution of the signed area of closed walks of length $N$ is given by 
%part in %the expansion of $H^{\bfn}$,
\begin{equation}
[u^0v^0] \big(u+u^{-1}+v+v^{-1}\big)^{\bfn}={\sum_{A=-\lfloor N^2/16 \rfloor}^{\lfloor N^2/16\rfloor} C_{\bfn}(A)\; \Q^{ A}},\label{nice}
\end{equation}
where $C_{\bfn}(A)$ counts the closed walks of length $\bfn$ enclosing a signed area~$A$.
For example, one easily checks that $[ u^0v^0] \big(u+u^{-1}+v+v^{-1}\big)^{4}=28 + 4 \Q+4\Q^{-1}$, indicating that among the $\binom{4}{2}^2=36$ closed walks making $4$ steps $C_{4}(0)=28$ enclose a signed area $A=0$ and $C_{4}(1)=C_{4}(-1)=4$ enclose a signed area $A=\pm 1$. 

\pagebreak

\section{The Hofstadter model}

{In any irreducible representation of the operator relation $vu=Quv$, the central element~$\Q$
will be represented by a number. Restricting to unitary representations, for which $u^\dagger = u^{-1}$,
$v^\dagger = v^{-1}$, $\Q$ will necessarily be {a complex number of norm unity, i.e., a phase}. %, $\Q=e^{i \phi}$. 
This provides a mapping
between the $u,v$ representation for walks and quantum mechanics, interpreting $u$ and $v$ as
unitary operators acting on a quantum Hilbert space, and the operator $u+u^{-1}+v+v^{-1}$ as the Hamiltonian
of a quantum system.

In fact, such a quantum system exists and corresponds to a well-known model in physics.
Interpreting $u$ and $v$ as operators that generate hops of a quantum particle by one link on the
square lattice, the %fundamental
 non-commutativity relation $vu=Quv$ indicates that translations
of the particle in the horizontal and vertical directions do not commute. This can be interpreted as that the particle is charged
and coupled to a homogeneous magnetic field perpendicular to the lattice.
The magnetic flux associated to this (constant) vector field is $\Phi a^2$ for any surface of area $a^2$.
 
Let us now take $\Q={\rme}^{{\rmi}2\pi\Phi/\Phi_o}$, where $\Phi$} is the magnetic flux through any unit lattice cell (i.e.~for the surface of the square of width $a=1$)
and where $\Phi_o=h/c$ is the flux quantum ($h$ is the Planck constant and $c$ the particle's charge)}. The Hermitian operator
\begin{equation} H= u+u^{-1}+v+v^{-1} \label{HHof}\end{equation} 
then becomes a Hamiltonian modelling
a quantum particle hopping on a square lattice and coupled to a perpendicular magnetic field. This
 model is known as the Hofstadter~model~\cite{Hof}.}

To make the physics connection completely explicit, we note that in quantum mechanics the
hopping operators $u$ and $v$ are written as
\begin{equation*}
u=\rme^{\rmi(p_x-c A_x)/\hbar}\text{\quad and \quad}
 v=\rme^{\rmi(p_y-c A_y)/\hbar}, \label{uv}
\end{equation*}
where $A_x=-\Phi y$\, and $A_y=0$ are the two components of the vector potential of the magnetic field
in the Landau gauge and $p_x=-\rmi\hbar {\partial_x}$ and $p_y=-\rmi\hbar {\partial_y}$ those of the
momentum operator (where we use the standard notation $\hbar=h/(2\pi)$). Thus, assuming~\eqref{uv}, the relation 
$vu=Quv$ follows from
the Baker--Campbell--Hausdorff formula, using the Heisenberg commutators
$[x,p_x]=[y,p_y]=\rmi\hbar$ (this identity is often called the ``canonical commutation relation'').

One now introduces the quantum state $\Psi_{m,n}$ representing the probability amplitude of the
particle being at lattice site $(m,n)$, on which hopping operators act as
\begin{equation*}
u \Psi_{m, n} = {\rme}^{{\rmi} c n \Phi/\hbar} \Psi _{m + 1, n} ~,\quad v \Psi_{m, n} = \Psi _{m , n +1}.
\end{equation*}
As $c \Phi/\hbar$ = $2\pi \Phi/(h/c) = 2\pi \Phi/\Phi_o$, the factor appearing in the action of $u$ on $\psi_{m,n}$ is $\Q^n$.
Using translation invariance in the horizontal direction we can further choose $\Psi_{m,n}$ to be an
eigenstate of $p_x$, that is, $\Psi_{m,n}={\rme}^{{\rmi}m k_x }\Phi_n$. The action of $u,v$ on $\Phi_n$ becomes
\begin{equation*}
u \Phi_n = {\rme}^{{\rmi} k_x} \Q^n \Phi_n ~,~~~ v \Phi_n = \Phi_{n+1}.
\end{equation*}
For the Hofstadter model, the
Schr\"odinger equation $H \Psi = E \Psi$ (which determines the eigenvalue $E$ of the spectrum) can be rewritten as
\begin{equation*}
(u+u^{-1}+v+v^{-1}) \Psi_{m, n}=E\Psi_{m, n}\Rightarrow \Phi_{n+1}+\Phi_{n-1}+(\Q^n {\rme}^{{\rmi} k_x}
+\Q^{-n} {\rme}^{-{\rmi}k_x})\Phi_n=E\Phi_n.
\end{equation*}

\pagebreak
Going a step further, a simplification arises when the flux is \textit{rational} (i.e.~when one has $\Q=\rme^{\rmi2\pi p/q}$ with $p,q$
two coprime integers): It induces a $q$-periodicity of the Schr\"odinger equation in the vertical direction.
% allowing us to choose Bloch states
%direction. Bloch's theorem then implies
Then, as Bloch's theorem states that solutions to the Schr\"odinger equation in a periodic potential take the form 
of a plane wave modulated by a periodic function ${\tilde \Phi}_n$, we can write
\be
\Phi_n = {\rme}^{{\rmi} n k_y } {\tilde \Phi}_n ~,~~~ {\tilde \Phi}_{n+q} = {\tilde \Phi}_n.
\label{Bloch}\ee
Indeed, for this rational flux $\Q^q =1$, thus $u^q , v^q$ become Casimirs (a physicist's term for central elements), and the
choice of Bloch states~\eqref{Bloch} can be interpreted mathematically as choosing an irreducible representation
of the $u,v$ algebra. Acting on such states, $u^q$ and $v^q$ become
$u^q = {\rme}^{{\rmi}q k_x} $ and $v^q = \rme^{{\rmi}q k_y}$.
One ends up with $u$ and $v$, acting on ${\tilde \Phi}_n$, becoming the $q \times q$ matrices
\bgroup
\setlength\arraycolsep{2pt}
\be u=\rme^{\rmi k_x}\begin{pmatrix}
\Q\quad & 0 & 0 & \cdots & 0 & 0\\
0\quad &\Q^2& \quad 0\quad & \cdots & 0 & 0 \\
0\quad & 0 & \Q^3 & \cdots & 0 & 0 \\
\vdots & \vdots & \vdots & \ddots & \vdots & \vdots \\
0\quad & 0 & 0 & \cdots & \Q^{q-1}& 0 \\
0\quad & 0 & 0 & \cdots & \quad 0\quad & 1\\
\end{pmatrix}
\text{\quad and \quad} v=\rme^{\rmi k_y}\begin{pmatrix}
0\quad & 1 & 0 & \cdots & 0 & 0\\
0\quad &0 & \quad 1\quad & \cdots & 0 & 0 \\
0\quad & 0 & 0 & \cdots & 0 & 0 \\
\vdots & \vdots & \vdots & \ddots & \vdots & \vdots \\
0\quad & 0 & 0 & \cdots & 0& 1\\
1\quad & 0 & 0 & \cdots & \quad 0\quad & 0 \\
\end{pmatrix}\nonumber\ee
\egroup
involving two real quantities $k_x$ and $k_y$.
Finding the energy spectrum (which depends on $k_x$\linebreak and $k_y$) reduces to computing
the eigenvalues $E_1,\ldots, E_q$ of the $q\times q$ Hamiltonian matrix 
$H_q:=u+u^{-1}+v+v^{-1}$, i.e.
\bgroup
\setlength\arraycolsep{2pt}
\be 
H_q=\begin{pmatrix}
\Q {\rme}^{{\rmi} k_x}+\Q^{-1} {\rme}^{-{\rmi}k_x} & {\rme}^{{\rmi}k_y} & 0 & \cdots & 0 & {\rme}^{-{\rmi}k_y}\\
{\rme}^{-{\rmi}k_y} &\Q^2 {\rme}^{{\rmi}k_x}+\Q^{-2}{{\rme}^{-{\rmi}k_x}} & \quad {\rme}^{{\rmi}k_y}\quad & \cdots & 0 & 0 \\
0 & {\rme}^{-{\rmi}k_y} & () & \cdots & 0 & 0 \\
\vdots & \vdots & \vdots & \ddots & \vdots & \vdots \\
0 & 0 & 0 & \cdots & () & {\rme}^{{\rmi}k_y} \\
{\rme}^{{\rmi}k_y} & 0 & 0 & \cdots & \quad {\rme}^{-{\rmi}k_y}\quad & \Q^q {\rme}^{{\rmi}k_x}+\Q^{-q}{{\rme}^{-{\rmi}k_x} } \\
\end{pmatrix}
\nonumber.\ee
\egroup

{All the machinery of quantum mechanics is now at our disposal. Selecting as in~\eqref{nice} the 
$u^0v^0$ monomial of $\big(u+u^{-1}+v+v^{-1}\big)^{\bfn}$
translates in the quantum world to computing the trace of $H_q^{{\bfn}}$. 
The quantum trace is defined as}
\be \textbf{Tr}\, H_{q}^{\bfn} {:= \frac{1}{q}\int_{-\pi}^{\pi}\int_{-\pi}^{\pi} \frac{d k_x}{2\pi}\frac{d k_y}{2\pi}\; {\rm tr}\, H_q^{\bfn}}
= \frac{1}{q}\int_{-\pi}^{\pi}\int_{-\pi}^{\pi} \frac{d k_x}{2\pi}\frac{d k_y}{2\pi} \sum_{i= 1}^{q} E_i^{\bfn}, %= \frac{1}{q}\int_{-\pi}^{\pi}\int_{-\pi}^{\pi} \frac{d k_x}{2\pi}\frac{d k_y}{2\pi} \tr H_q^{\bfn}
\label{qtrace}\ee
that is, one sums over the $q$ eigenvalues $E_i$ of $H_q$ (yielding the standard matrix trace ${\rm tr}\,H_q^{\bfn}$) and integrates over $k_x$ and $k_y$ while enforcing
a continuous normalization in $k_x ,k_y$ and
one rescales by a factor $1/q$
(thus, if one considers for example the $q\times q$ identity matrix~$I_q$,
one has ${\rm tr}\, I_{q}=q$, while $\textbf{Tr}\, I_{q}=1$).
 Under this definition of the trace,
$\textbf{Tr}\,u^m v^n = \delta_{m,0}\delta_{n,0}$ (integration over $k_x,k_y$ eliminates the traces of
terms involving $u^{qm}$ and $v^{qn}$). We thus get a first noteworthy result (also obtained via another approach by Bellisard et al.~in~\cite{BellissardCamachoBarelliClaro1997}):
\begin{theorem}
Assuming $q>\frac{N^2}{8}$,  the signed area enumeration of closed paths is given by
\be
\sum_{A} C_{\bfn}(A)\; \Q^{ A}
={\normalfont\textbf{Tr}}\; H_q^{{\bfn}}
\label{eureka}.
\ee
\end{theorem}
%To isolate the coefficients $C_{\bfn} (A)$ the integer $q$ has to be taken as a free parameter, allowing for an expansion
%of the trace in the $\Q^A$ basis so that the enumeration~\eqref{eureka} can take place.
\pagebreak

\section{The signed area enumeration}
%{\red since $\Q^q = 1$ and in this case $\textbf{Tr}\, u^m v^n = \sum_{s,t=-\infty}^\infty \delta_{m+sq,n+tq}$.}

It is known (see e.g.~William Chambers' book~\cite{Chambers}) that the determinant of the matrix $I_q-zH_q$ satisfies
\be\nonumber\label{ZZn} \det(I_q-z H_q)=\left(\sum_{n=0}^{\lfloor q/2 \rfloor} (-1)^nZ(n)z^{2n}\right)-2\big(\cos(q k_x)+\cos(q k_y)\big)z^q,
\ee
where the $Z(n)$'s are independent of $k_x$ and $k_y$ and $Z(0)=1$.
Christian Kreft~\cite{Kreft} was able to rewrite~$Z(n)$ in a closed form as trigonometric
multiple nested sums
\be\label{Zn} Z(n) = \sum_{k_1=1}^{q-2n+2} \sum_{k_2=1}^{k_1} \cdots \sum_{k_{n}=1}^{k_{n-1}} \prod_{j=1}^n s_{k_j+2n-2j},\ee
\noindent where
\be s_{k}=4\sin^2(\pi k p/q)=2-Q^k-Q^{-k}\label{sk}.\ee
We call $s_k$ the spectral function of the model. This is the starting point for the signed area enumeration.
We give here a summary of the procedure, more details can be found in~\cite{1,2}.
First introduce the coefficients $b(n)$ via
\be
-\log\left(\sum_{n=0}^{\lfloor q/2 \rfloor}(-1)^n Z(n) z^{2n} \right)=\sum_{n=1}^{\infty} b(n) z^{2n}.\label{bn}
\ee
The $b(n)$ are related to the desired traces. Start by noting that
\be\nonumber
\Tr H_q^{2n} = \frac{1}{q} \tr H_q^{2n} ~~~\text{for}~~~ n<q. % \qquad \text{(recall that $N=2n$ is the length of the walk).}
\ee
Indeed, one has $\tr u^i v^j = q\; \delta_{i,0}\delta_{j,0}$ for $i,j <q$ and so the values of the
Casimirs $k_x,k_y$ do not appear, making the integration over them in~\eqref{qtrace} trivial. Then the identity
\be\nonumber
-\log\,\det(I_q -z H_q)=-\tr \log(I_q-z H_q) = \sum_{n=1}^\infty \frac{z^n}{n} \tr H_q^n=\sum_{n=1}^{\lfloor q/2\rfloor}  b(n) z^{2n} +O(z^q)
\ee
implies that the quantum trace~\eqref{qtrace} is proportional to $b(n)$ for $n<q$
\be \nonumber \textbf{ Tr}\, H_q^{2n}= \frac{2n}{q} b(n).
\ee
%Note that $ \tr H_q^n=0$  for odd $n$ is combinatorially obvious: there are no closed walks of odd lengths.
Now, by keeping $q$ as a {free} parameter and extending it to arbitrarily big values, the quantum trace can be calculated for all $n$.

Note that the term of order $z^n$ in $\det(I_q -z H_q)$ is $-(z^n/n) \tr H_q^n$ plus terms involving
products of traces, $\tr H_q^{n_1}\, \tr H_q^{n_2} \cdots$ with $n_1 + n_2 + \cdots = n$. As each trace 
$\tr H_q^{n_i}$ contributes an overall factor $q$, $b(n)$ can also be obtained as the order $q$
term in $Z(n)$, ignoring terms of higher order $q^2,\dots,q^n$. Since each sum in Formula~\eqref{Zn}
contributes a factor of $q$, we get the following explicit expression for $b(n)$: 
\be b(n)= \sum_{j=1}^n {\sum_{\substack{l_1, l_2, \ldots, l_{j} \\ { \rm composition}\;{\rm of}\; n }} \hskip -0.4cm 
{c(l_1,l_2,\ldots,l_{j} )} }{\sum _{k=1}^{q-j+1} s^{l_{j}}_{k+j-1}\cdots s^{l_2}_{k+1} {s}^{l_1}_k},
\label{sonice}\ee
\be \text{where\ } \nonumber c(l_1,l_2,\ldots,l_{j}):= \frac{\binom{l_1+l_2}{l_1}}{l_1+l_2}\;\; l_2\frac{\binom{l_2+l_3}{l_2}}{l_2+l_3}\;\cdots \;\; l_{{j}-1}\frac{\binom{l_{{j}-1}+l_{j}}{l_{{j}-1}}}{l_{{j}-1}+l_{j}}.
\ee

Here, the coefficients $c(l_1,l_2,\ldots,l_{j})$ are labelled by the compositions $l_1,l_2,\ldots,l_{j}$ of~$n$, 
i.e.~the ordered partitions of $n$ (thus, there are $2^{n-1}$ compositions of $n$, for example $3=3,2+1,1+2,1+1+1$).
Note that the expression for $c(l_1,l_2,\ldots,l_{j} )$ is closely related to {the enumeration
of Dyck paths} (up to a factor~$l_1$); see~Christian Krattenthaler's article~\cite[p.~516]{krat}. We have further elaborated on this relation in~\cite{recent}.

%As stated, only single trigonometric sums appear in~\eqref{sonice}.
%, weighted by $c(l_1,l_2,\ldots,l_{j} )$ and summed over all the compositions of $n$.
% ${\sum _{k=1}^{\Q-j+1} s^{l_{j}}_{k+j-1}\cdots s^{l_2}_{k+1} {s}^{l_1}_k}$. 
Putting everything together, we get\vspace{-2mm}
\be \nonumber \textbf{ Tr}\; H_q^{2n}= \sum_{j=1}^n {\sum_{\substack{l_1, l_2, \ldots, l_{j} \\ { \rm composition}\;{\rm of}\;n}} \hskip -0.4cm 
{c(l_1,l_2,\ldots,l_{j} )}}\;\frac{2n}{q} \ \ {\sum _{k=1}^{q-j+1} s^{l_{j}}_{k+j-1}\cdots s^{l_2}_{k+1} {s}^{l_1}_k}.
\ee
What is more, the trigonometric sums involved in this formula can be computed, keeping~$q$ as a free parameter. % as mentioned before. 
Finally, returning to~\eqref{eureka} (with $N=2n$) the desired number of closed walks of given area is given by the following theorem.
\begin{theorem}\label{coco}
The number of closed walks of length $2n$ enclosing a given signed~area~$A$~is\vspace{-1.8mm}
		\begin{align}\nonumber C_{2n}(A) &= \ 2 n \times \sum_{j=1}^{n} {\sum_{\substack{
		l_1, l_2, \ldots, l_j\\ \text{ composition of}\; n}} {\frac{\binom{l_1+l_2}{l_1}}{l_1+l_2}\;\; l_2\frac{\binom{l_2+l_3}{l_2}}{l_2+l_3}\;\cdots \;\; l_{{j}-1}\frac{\binom{l_{{j}-1}+l_{j}}{l_{{j}-1}}}{l_{{j}-1}+l_{j}}}}~ \times\\&\nonumber \hskip -1.6cm{\sum_{k_3= 0}^{2l_3 
 }\sum_{k_4= 0}^{ 2l_4}\ldots \sum_{k_{j}= 0}^{2l_{j}}\prod_{i=3}^{j}\binom{2l_i}{k_i}} {\binom{2l_1}{l_1 +A+\sum_{i=3}^{j}(i-2)(k_i-l_i)}\binom{2l_2}{l_2 -A-\sum_{i=3}^{j}(i-1)(k_i-l_i)}}.%\\&\label{coco}
 \end{align} 
\end{theorem}
This formula grows quickly in complexity since one has to sum over $2^{n}-1$ compositions.
 Its complexity is analysed in more detail in the following proposition.

\begin{proposition}
The formula for $C_{2n}(A)$ in Theorem~\ref{coco} involves asymptotically, up to some polynomial factor,
$(4/(5-\sqrt{17}))^n \approx 4.56^{n}$ summands.
\end{proposition}
\begin{proof}
This number of summands is given by
\begin{equation*}
r(n):=\sum_{j=1}^{n} \sum_{\substack{l_1, l_2, \ldots, l_j\\ \text{ composition of}\; n}} (2l_3+1) (2l_4+1) \dotsm (2l_j+1).
\end{equation*}
The sequence starts like $(r(n))_{n\geq 1} =(1,2,6,24,106,480,2186,9968,45466,\dots)$.
Thus, $r(n)$ counts compositions of $n$ where each summand $l_i$ (for $i\geq 3$) can have $2 l_i+1$ colours.
Let us consider first compositions of $n$ where each summand $l_i$ (for $i\geq 1$) can have $2 l_i+1$ colours;
their generating function is 
\begin{equation*} S(z)= \frac{1}{1-\sum_{i\geq 1} (2 i+1) z^i} = \frac { \left( z-1 \right) ^2}{2 z^2-5 z+1}.\end{equation*}
This corresponds to the sequence $\oeis{A060801}$ in the On-line Encyclopedia of Integer Sequences.
In our case, the $i\geq 3$ constraint modifies a little bit the generating function and one has
to sum the compositions having 1 or 2 parts and those having 3 parts or more; one gets the following generating function
\begin{align*}R(z)&= \sum_{n\geq 1} r_n z^n = \frac{z}{1-z}+ \left(\frac{z}{1-z}\right)^2 + \left(\frac{z}{1-z}\right)^2 (S(z)-1)\\
&= \frac {z^2-4z+1}{2z^2-5z+1} \frac{z}{1-z}. 
\end{align*}
It entails $r(n) =5 r(n-1) - 2 r(n-2)-2$ (with $r(1)=1$ and $r(2)=2$); accordingly, $r(n)$ grows like~$\phi^n$ with~$\phi=4/(5-\sqrt{17})$,
which is coherent with the fact that $1/\phi$ is the dominant pole of $R(z)$.
%the entry \oeis{A060801} in the On-line Encyclopedia of Integer Sequences). 
In conclusion, our formula for $C_{2n}(A)$ involves in total asymptotically $\phi^{n}\approx 4.56^{n}$ summands (each leading additionally to a polynomial cost in $n$ for the product of all the binomials), which is still much less\footnote{We thank one of the referees for drawing our attention to this point.} 
 than the naive generation of all $\binom{2n}{n}^2$ closed walks of length $2n$, which would be~of~cost~$>16^n/(\pi n)$. 
\end{proof}

Note that it is in fact possible to compute $C_{2n}$ in polynomial time:
 Using the non-commutative relations~\eqref{add_rel},
the expansion of $(u+v+u^{-1}+v^{-1})^{2n}$ simplifies a lot and 
in fact has $O(n^4)$ monomials $u^i v^j Q^k$. 
This gives an algorithm of complexity $O(n^6)$ to compute $C_{2n}(A)$.
Thus, our formula (of exponential cost) in Theorem~\ref{coco} 
is not the fastest way to compute $C_{2n}(A)$, 
but it has the benefit of being the first explicit formula (as far as we are aware of!). 

\smallskip

Let us end this section with a probabilistic remark.
In the limit of the elementary lattice size $a \to 0$ and the walk length $2n \to \infty$ with the
scaling $n a^2=t$, walks converge to Brownian motion curves and we recover the continuum limit of
a particle moving on the plane in a constant magnetic field. To implement this limit,
we rescale the %elementary plaquette
lattice cell area to $a^2$, which amounts to setting $A \to A/a^2$ in $C_{2n} (A)$. 
Numerical simulations then suggest the following conjecture.

\begin{conjecture}The signed area of closed walks of length $2n$ converges, after rescaling, to the following distribution
(for any $\alpha>0$)
\be \nonumber \frac{2n\;C_{2n}(\alpha n )} {\binom{2n}{n}^2}\to \frac{\pi}{\cosh^2(\alpha \pi)}.
\ee
\end{conjecture}

\begin{figure}[h]
\includegraphics[height=5cm]{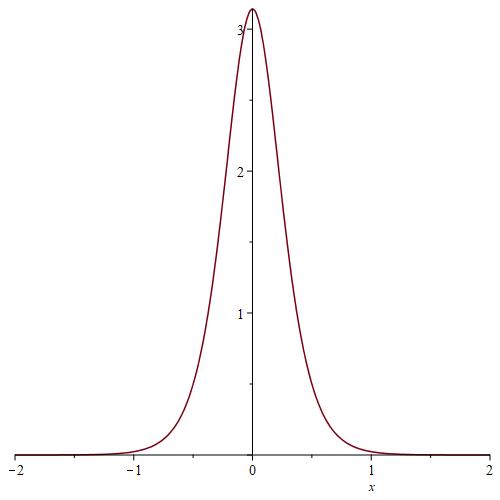}
\setlength\abovecaptionskip{-2mm}
\setlength\belowcaptionskip{-1mm}
\caption{We conjecture that the distribution of the signed area asymptotically follows\dots  not a Gaussian limit law (as it may be thought at first glance), 
but the $1/\cosh^2$ distribution of Paul L\'evy. 
For small values of $A$, and $n$ up to~$70$, we checked numerically that the convergence gets better when $n$ increases.}
\end{figure}

This conjecture is consistent with the law for the distribution of the signed area enclosed by a Brownian curve after a time $t$
(obtained by Paul L\'evy in 1950; see~\cite{Levy1950,KhandekarWiegel1988}). %Levy1965
It can also be obtained directly in the continuum limit by considering
the partition function of a quantum particle in a magnetic field with a Landau level energy spectrum.
 
\pagebreak

\section{Exclusion statistics}
	
%{Why in (\ref{Zn}) and (\ref{bn}) the particular choice of the notations {\red $Z(n)$} and {\red $b(n)$}? In statistical mechanics $Z(n)$ usually refers to an $n$-body partition function and $b(n)$ to its associated $n$-{th} cluster coefficient.
{The quantities $Z(n)$ and $b(n)$ introduced previously admit a statistical mechanical interpretation.
Let us write the spectral function} $s_k$ in~\eqref{sk} as $s_k={\rme}^{-\beta\epsilon_k}$ ($\beta$
is the inverse temperature) {and interpret is
as the Boltzmann factor} for a quantum 1-body spectrum $\epsilon_k$ labelled by an integer $k$. The structure of $Z(n)$ in~\eqref{Zn} then precisely corresponds to an $n$-body partition function for a gas of
particles {with 1-body spectrum $\epsilon_k$ and exclusion statistics $g=2$: The $+2$ shifts
in the spectral function arguments ensure
that no two particles can occupy adjacent quantum states.} Exclusion statistics is, again, a purely quantum concept which describes the statistical mechanical properties of identical particles.
{Ordinary} particles are either bosons ($g=0$), {which can occupy the same quantum state,} or fermions ($g=1$), {which cannot occupy the same quantum state. We see that square-lattice walks map to systems
with statistics beyond Fermi exclusion, {in which particles can occupy neither the same state nor
adjacent states. In a sense, each particle excludes two quantum states, thus $g=2$}. In general, for $g$-exclusion
{particles} the $n$-body partition function~\eqref{Zn} would become
\be
 Z(n) 
 = \sum_{k_1=1}^{q-gn+g} \sum_{k_2=1}^{k_1} \cdots \sum_{k_{n}=1}^{k_{n-1}}
s_{k_1+gn-g}
s_{k_2+gn-2g} \cdots s_{k_{n-1}+g}
s_{k_{n}},
\label{Zqg}\ee
where one observes a shift {$g$ instead of $2$} in the arguments of the spectral function. In line with~\eqref{bn},~\eqref{sonice}
the associated $n$-th cluster coefficient {can be shown to take the form}
\be b(n)=\sum_{j=1}^n \sum_{\substack{l_1, l_2, \ldots, l_{j} \\ { g-{\rm composition}\;{\rm of}\;n}} }\hskip -0.4cm 
{c_g(l_1,l_2,\ldots,l_{j} )} {\sum _{k=1}^{q-j+1} s^{l_{j}}_{k+j-1}\cdots s^{l_2}_{k+1} {s}^{l_1}_k},
\label{suf}\ee
where
\be \nonumber {c_g (l_1,l_2,\ldots,l_{j})} := 
{{{\frac{(l_1+\dots +l_{g-1}-1)!}{l_1! \cdots l_{g-1}!}}~
\prod_{i=1}^{j-g+1} \binom{l_i+\dots +l_{i+g-1}-1 }{l_{i+g-1}}}}.
\ee	
In~\eqref{suf} one sums over all $g$-compositions of the integer $n$, obtained by inserting at will inside the usual
compositions (i.e., the $2$-compositions) no more than $g-2$ zeroes in succession.
For example, for $n=3$ and $g=3$ one has $9$ such $3$-compositions: 
\begin{equation*}
 3,2+1,1+2,1+1+1,2+0+1,1+0+2,1+0+1+1,1+1+0+1,1+0+1+0+1.
\end{equation*}
For general $g$ there are $g^{n-1}$ such {$g$-compositions} of the integer $n$ (see~\cite{brian} for an analysis of these extended compositions, also called multicompositions).
		
One has reached the conclusion that the signed area enumeration for walks on the square lattice is described
by a quantum gas of particles with statistical exclusion $g=2$. {To relate this explicitly to properties
of} the Hofstadter Hamiltonian itself, let us perform on the hopping lattice operators $u$ and $v$ the %modular 
transformation 
\be \nonumber
 u \to - u\; v,~~~ v \to v,
\ee
which leave their own commutation relation invariant 
to get the new Hamiltonian
\be\label{new} H=-u\;v -v^{-1} u^{-1}+v+v^{-1}\ee
still describing the same walks but on a deformed lattice.
%\begin{figure}[h]
%\centering
%\includegraphics[width=\textwidth]{untitled}
%\begin{center}
%\hspace{-1cm}
%\vspace{4.8cm}
%\includegraphics[width=1.3\textwidth]{solong}
%
%%%%%\vskip -5cm
%%%%%\includegraphics[scale=1.2]{Untitled}
%%%%%\vskip -15cm
%
%\caption{\red suppress ??: The deformed square lattice after the modular transformation.}
%\end{center}
%\end{figure}

This new Hamiltonian (if one compares it with the initial Hamiltonian operator~\eqref{HHof} of the Hofstadter model)
has the advantage to lead to simpler matrix
%) We now consider the following matrix associated to the Hamiltonian~\eqref{new} 
\begin{equation} I_q-z H_q =\begin{pmatrix}
{1}& -{(1-\Q)}z & 0 & \cdots & 0 & 0 \\
-{(1-{\frac{1}{\Q}})}z & {1} &- {(1-\Q^2)}z & \cdots& 0 & 0 \\
0 &- (1-{\frac{1}{\Q^{2}}})z & {1} &\cdots & 0 & 0 \\
\vdots & \vdots & \vdots & \ddots & \vdots & \vdots \\
0 & 0 & 0 & \cdots & {1} & -{(1-{\Q^{q-1}})}z \\
0 & 0 & 0 & \cdots & -{(1-{\frac{1}{\Q^{q-1}}})}z& {1} \\
\end{pmatrix},\label{cocobis}\end{equation}
where we set $k_x=k_y=0$ for simplicity (since, as we explained in Section 3, $k_x$ and $k_y$ do not appear in the counting formula). 
The Hofstadter spectral function~\eqref{sk} {becomes}
\be \nonumber
s_k= (1-\Q^k )(1-{\frac{1}{\Q^{k}}} ).
\ee
{The matrix~\eqref{cocobis} is a particular case of the more general class of matrices having the following shape\footnote{We hope that the reader 
will easily distinguish between too similar (but unrelated!) notations: the \textit{function} $g(k)$ for the entries of the matrix $I_q-zH_q$, and the \textit{integer} $g$ (the parameter of the exclusion model, a standard notation in the literature).}
\be 
I_q-zH_q =\begin{pmatrix}
{1}& -{f(1)}z & {0} & \cdots & 0 & -{g(q)}z \\
-{g(1)}z & {1} & -{f(2)}z & \cdots& 0 & 0 \\
0 & -{g(2)}z & {1} &\cdots & 0 & 0 \\
\vdots & \vdots & \vdots & {\ddots} & \vdots & \vdots \\
0 & 0 & 0 & \cdots & {1} & -{f(q-1)}z \\
-{f(q)}z & 0 & 0 & \cdots & -{g(q-1)}z & {1} \\
\end{pmatrix}\label{ca}\ee
and associated spectral functions 
\be \nonumber
{s_k= f(k)g(k)},
\ee
which become the building blocks of the $Z(n)$'s in~\eqref{Zn} (up to spurious {``umklapp'' terms,
a name deriving from momentum periodicity effects on lattice quantum models,}
which disappear if $f(q)$ and $g(q)$ both vanish). }}

For statistics $g=3$, the matrix~\eqref{ca} generalizes in a natural way to

\hspace*{-1mm}\resizebox{.982\linewidth}{!}{\hspace*{-8mm} \begin{minipage}{1.10\linewidth}%\hspace*{-10mm} 
\be \hspace*{2mm} 
I_q-zH_q=\begin{pmatrix}
{1}& -{f(1)}z & 0 &0 & \cdots & 0 &-{g(q-1)} z &{0} \\
{0} & {1} & -{f(2)}z &0&\cdots & 0&0& -{g(q)} z \\
 -{g(1)} z & {0} & {1} &-{f(3)}z &\cdots &0 & 0&0 \\
 0&-{g(2)}z&{0}&{1}&\cdots &0 & 0&0\\
 \vdots&\vdots & \vdots &\vdots &{\ddots} &\vdots &\vdots & \vdots\\
0 &0 & 0 &0& \cdots &{1} & -{f(q-2)}z&0 \\
0 & 0 & 0 &0&\cdots& {0}&{1} &-{f(q-1)}z \\
-{f(q)}z & 0 & 0 &0 & \cdots&-{g(q-2)} z&{0} & {1} \\
\end{pmatrix},
\label{mamatrix}\ee
\end{minipage}
}\newline
that is, with an extra vanishing paradiagonal below the unity main diagonal, which is the manifestation
of the stronger $g=3$ exclusion. 
\pagebreak

\noindent The spectral function corresponding to this matrix is
\begin{equation*}
s_k= g(k)f(k)f(k+1)
\end{equation*}
and the determinant $\det(I_q -z H_q)$ assumes the form~\eqref{Zqg} with $g=3$.
For $g$-exclusion the generalization of ~\eqref{mamatrix} amounts to a Hamiltonian of the form
\be \label{newbis}H=F(u)v+v^{{1-g}}G(u),
\ee
and $I-zH$ is then a matrix with $g-2$ vanishing paradiagonals below the main diagonal (here $q$ is left arbitrary but always understood to be larger than $g$). The spectral parameters of this matrix are}
\be \nonumber
F(\Q^k)=f(k),\quad G(\Q^k)=g(k),
\ee
and the spectral function is
\be \nonumber
{s_k= g(k)f(k)f(k+1)\ldots f(k+g-2)}.
\ee 
Clearly the Hofstadter Hamiltonian~\eqref{new}, which rewrites as 
$H=(1-u)v+v^{{1-2}}(1-u^{-1})$,
 is a particular case of~\eqref{newbis} with $g=2$ and $F(u)=1-u$, $G(u)=1-u^{-1}$.

\section{Chiral walks on the triangular lattice}
\begin{figure}[h]
\centering
\includegraphics[scale=0.18]{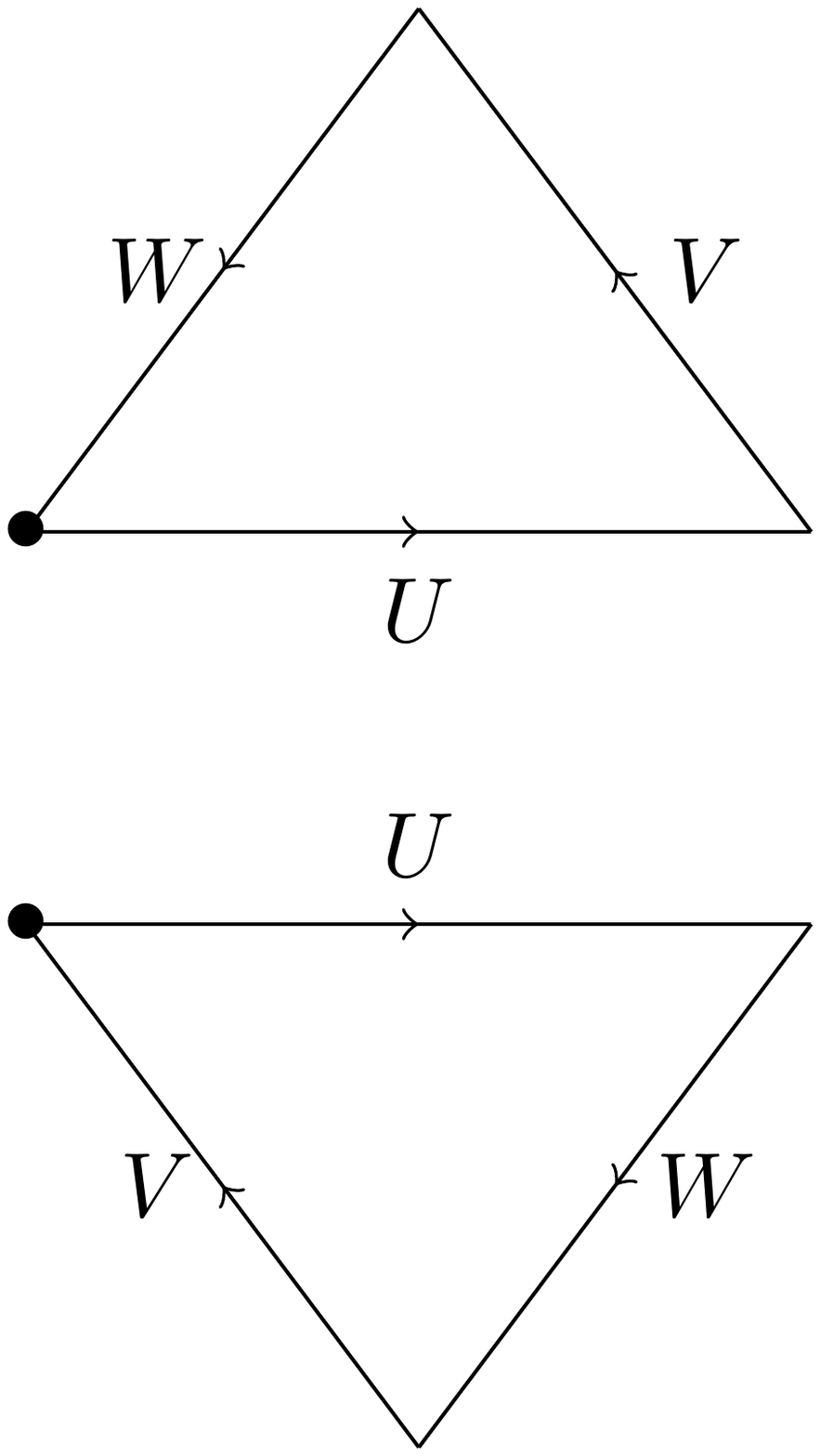}
\caption{The three hopping operators $U, V$ and $W$ on the triangular lattice.}\label{fig2}
\end{figure}

Let us illustrate this mechanism in the case of $g=3$ exclusion with the specific example of chiral walks on a triangular lattice. The three hopping operators $U, V$ and
 $W=\Q U^{-1} V^{-1}$ described in Figure~\ref{fig2} are such that $VU=\Q^2 UV$.

\begin{figure}[h]
\centering
\includegraphics[scale=0.26]{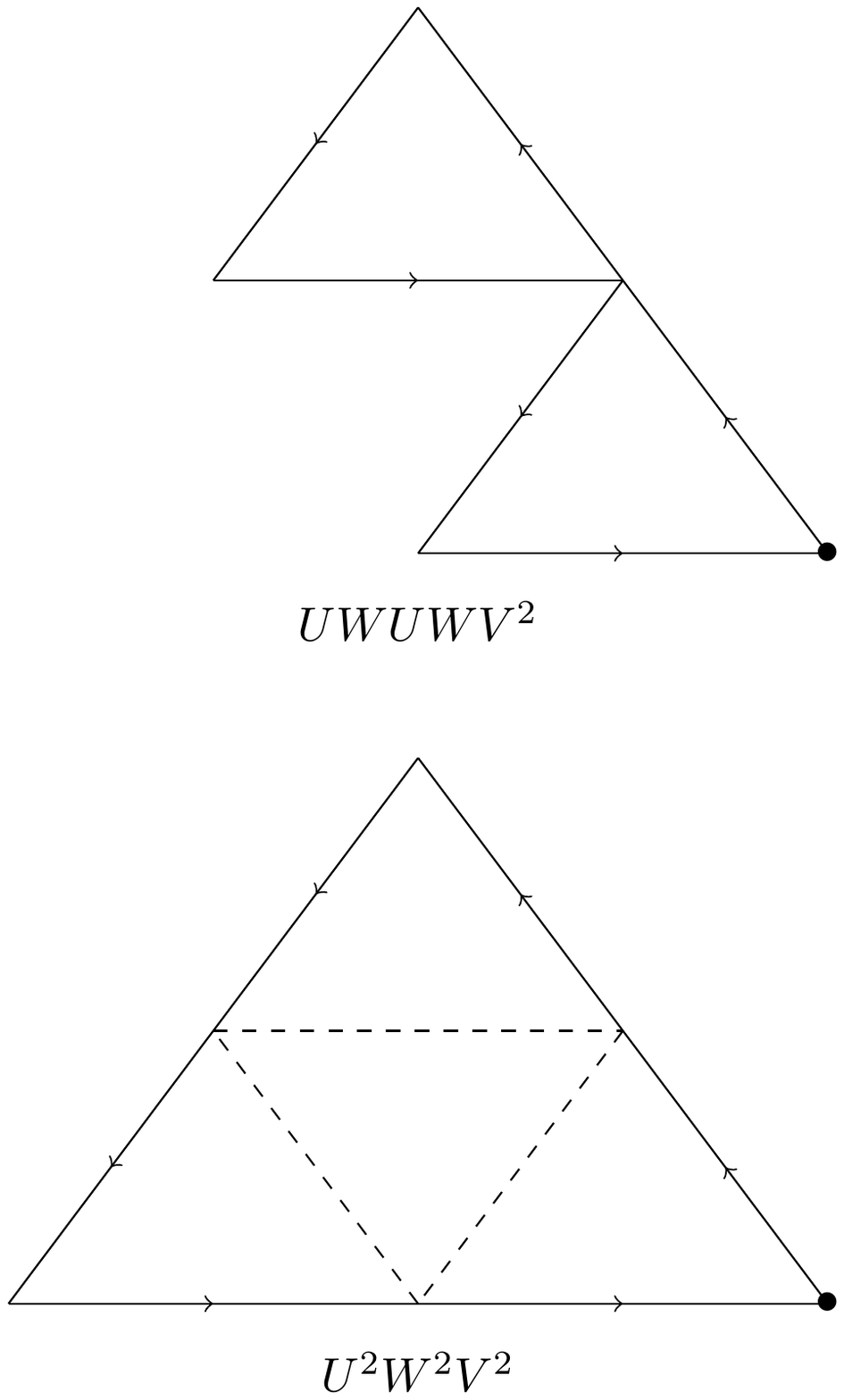}
\caption{Examples of closed chiral walks on the triangular lattice.}\label{fig3}
\end{figure}

The triangular lattice Hamiltonian is 
\be H=U+V+ W.\nonumber\label{triangle}\ee
It generates walks composed of triangles either pointing up and winding in the counterclockwise direction,
or pointing down and winding in the negative direction. In this sense, the walks are chiral.
The factor $\Q$ in the definition of $W$ and the factor $\Q^2$ in the commutation of $U,V$ are chosen
so that up-pointing (positive) triangles are assigned area $+1$. Figure~\ref{fig3} depicts some examples of chiral
walks on the triangular lattice.
\pagebreak

To bring $H$ %(\ref{triangle}) 
 to the exclusion form~\eqref{newbis} one chooses the representation
$U = -{\rmi} \, u \, v$ and $V = {\rmi}\, u^{-1} \, v$, with $u$ and $v$ as before,
in which case $H$ rewrites as
\be\nonumber
H = {\rmi}(- u+ u^{-1}) v + v^{-2}.
\ee
In this form, $H$ is indeed a Hamiltonian of the type~\eqref{newbis} for $g=3$ exclusion, with
$F(u) = {\rmi}(- u + u^{-1})$, $ G(u) = 1$, spectral parameters
\be\nonumber
f(k)= -{\rmi}(\Q^k-\frac{1}{\Q^k}), \quad g(k)= 1,
\ee
spectral function
\be
s_k = g(k)f(k)f(k+1)=4\sin (2 \pi p k / q)
\sin\big(2\pi p(k+1)/q\big),\label{OK}
\ee
and matrix 

\begin{center}
\resizebox{0.99\linewidth}{!}{
\begin{minipage}{177mm}
\setlength\arraycolsep{3pt}
\be\hspace*{0mm} I_q-zH_q=\begin{pmatrix}
{1}& \rmi(\Q-{\frac{1}{\Q}})z & 0 &0 & \cdots & 0 &-z &{0} \\
{0} & {1} & \rmi(\Q^2-{\frac{1}{\Q^{2}}})z &0&\cdots & 0&0& - z \\
 - z & {0} & {1} & \rmi(\Q^3-\frac{1}{\Q^{3}})z &\cdots &0 & 0&0 \\
 0&-z&{0}&{1}&\cdots &0 & 0&0\\
 \vdots&\vdots & \vdots &\vdots &{\ddots} &\vdots &\vdots & \vdots\\
0 &0 & 0 &0& \cdots &{1} & \rmi(\Q^{q-2}-{\frac{1}{\Q^{q-2}}}) z&0 \\
0 & 0 & 0 &0&\cdots& {0}&{1} & \rmi(\Q^{q-1}-{\frac{1}{\Q^{q-1}}})z \\
0 & 0 & 0 &0 & \cdots&- z&{0} & {1} \\
\end{pmatrix},\nonumber\ee
\end{minipage}}
\end{center}
which is of the type~\eqref{mamatrix} with a vanishing bottom-left entry.
The non-Hermiticity of the triangular Hamiltonian, and thus of $I_q-zH_q$, is a consequence of
the chiral nature of the walks.
					 	
The triangular signed area enumeration follows~\cite{2}, yielding an expression similar to Theorem~\ref{coco}
with the trigonometric single sums appearing in~\eqref{suf} involving the triangular spectral function
\eqref{OK} and the sum done over all {3}-compositions of the length of the triangular walks.

\section{Conclusion}

In conclusion, we have shown how tools from quantum and statistical physics allow for an
explicit enumeration of closed walks of fixed length and signed area on planar lattices.

The enumeration formulae rely on an explicit sum over compositions, and their number of terms grows
quickly with the length of the walk (although much less quickly than a brute-force counting formula).
It would certainly be rewarding to rewrite it as a sum with a smaller number of terms. The use of symmetry on the lattice or alternative ways to write the generator of walks (Hamiltonian)
may offer promise towards this goal. We leave this issue as well as other questions of interest to the lattice walk combinatorics community. 
\vspace*{-.5mm}
%%%%%%%%%%%%%%%%%%END%%%%%%%%%%%

\subsection*{Acknowledgments}
We thank the referees for their useful comments and suggestions, and Li Gan for a careful reading of the manuscript.
%\smallskip
\vspace*{-1.7mm}

\subsection*{Funding} 
A.P. thanks the National Science Foundation for its support under grant {\small NSF-PHY-2112729} and {\small PSC-CUNY} for its support under grants {\small 65109-00 53} and {\small 6D136-00~03}.
\vspace*{-3.1mm}

\bibliographystyle{SLC} 
\bibliography{OuvryPolychronakos}

\end{document}